\documentclass[%
 aip,
 apl,%
 amsmath,amssymb,
 preprint,%
]{revtex4-1}

\usepackage{graphicx}
\usepackage{dcolumn}
\usepackage{bm}
\usepackage{color}
\usepackage{textcomp}

\newif\ifcmnt
\cmnttrue

\ifcmnt
    \providecommand{\aucmnt}[1]{#1}
\else
    \providecommand{\aucmnt}[1]{}
\fi

\begin{document}


\title[Sample title]{Maximizing waveguide integration density with multi-plane photonics}

\author{Jeff Chiles}
\email{jeffrey.chiles@nist.gov}
\affiliation{ 
	National Institute of Standards and Technology\\Boulder, CO, USA
}%

\author{Sonia Buckley}%
\affiliation{ 
	National Institute of Standards and Technology\\Boulder, CO, USA
}%

\author{Nima Nader}%
\affiliation{ 
	National Institute of Standards and Technology\\Boulder, CO, USA
}%

\author{Sae Woo Nam}
\affiliation{ 
	National Institute of Standards and Technology\\Boulder, CO, USA
}%

\author{Richard P. Mirin}
\affiliation{ 
	National Institute of Standards and Technology\\Boulder, CO, USA
}%

\author{Jeffrey M. Shainline}
\affiliation{ 
	National Institute of Standards and Technology\\Boulder, CO, USA
}%

\date{\today}

\begin{abstract}
We propose and experimentally demonstrate a photonic routing architecture that can efficiently utilize the space of multi-plane (3D) photonic integration.  A wafer with three planes of amorphous silicon waveguides was fabricated and characterized, demonstrating $<3\times10^{-4}$ dB loss per out-of-plane waveguide crossing, $0.05 \pm 0.02 $ dB per interplane coupler, and microring resonators on three planes with a quality factors up to $8.2 \times 10^{4}$.  We also explore a phase velocity mapping strategy to mitigate the crosstalk between co-propagating waveguides on different planes.  These results expand the utility of 3D photonic integration for applications such as optical interconnects, neuromorphic computing and optical phased arrays.

\end{abstract}

\maketitle

\section{\label{sec:intro}INTRODUCTION}

One of the most prominent advantages of photonic integration is the ease with which signals can be routed over a wide range of distances without incurring excessive power penalties, losses, or crosstalk.  Photonic interconnects are a promising approach for applications requiring massive connectivity, such as phased arrays \cite{Sun2013,Abediasl2015} and optical transceivers \cite{Zhang2016}.  More recently, the field  of neuromorphic computing utilizing photonics \cite{tana20142,Shen2017,shainlinepra} has emerged as a research direction, motivated by the potential to realize all-to-all connectivity at the scale of $10^{3}$ synaptic connections per neuron \cite{shainlinepra}.  The footprint of the interconnections is minimized if signals can cross paths at least a similar number of times. For single-plane photonics, compact multimode waveguide crossings \cite{shainlinecrossing} with 0.02 dB loss per crossing have been demonstrated\cite{Zhang13}, allowing several dozen such junctions in a path without significantly impacting the power budget.  However, to achieve connectivity orders of magnitude greater, multi-planar (3D) photonic integration becomes necessary to minimize the crossing loss and to increase the maximum photonic waveguide density. \cite{Kang2013,Sacher2015,Hosseinnia2015,Yoo2016}.

Once the decision to expand vertically has been made, we are faced with many more choices concerning the platform: waveguide materials, confinement strength, interplane pitch, and interplane coupler (IPC) mechanism.  These elements are intricately related through their impact on the critical metrics of crossing loss, crosstalk, and the horizontal and vertical waveguide density that can be attained.  To minimize the crossing loss and crosstalk between out-of-plane waveguides, the optical modes must be sufficiently far apart to avoid scattering or evanescent coupling.  However, increasing the interplane pitch also compromises size and efficiency of the IPCs.  Previous work has demonstrated a two-plane crystalline/amorphous (c-Si/a-Si) platform with a 1.12 $\mu $m interplane pitch.  Such a large separation allows reasonable mitigation of crosstalk and crossing loss.  However, it also poses a challenge for the IPC, which suffered from  high loss (0.49 dB) and large dimensions ($\sim 200$ $\mu$m length)\cite{Itoh2016}.  To avoid these penalties, smaller pitches and weaker modal confinement can be pursued instead.  A silicon-nitride two-plane platform with a pitch of 900 nm was bridged with a 100 $\mu $m long adiabatic taper with $<$ 0.01 dB loss per coupler\cite{Shang2015}.  However, a consequence of the reduced inter-plane isolation was a severe penalty of 0.167 dB loss per out-of-plane waveguide crossing.  With even smaller gaps, considerably shorter couplers can be achieved with similar loss performance \cite{Hosseinnia2015}, but nothing is done to address the issues of crosstalk and crossing loss.  One way to circumvent these issues is to employ an additional intermediate routing plane to allow efficient coupling between smaller gaps, while maintaining a large separation in crossing areas; this has been realized with $3.1\times10^{-3}$ dB per crossing while co-integrating modulators and detectors on the same platform, showcasing the utility of 3D integration for high-density interconnect and transceiver applications \cite{sacher2016multilayer}.  However, the need to utilize an entire plane to augment the interplane pitch is a significant drawback to such an approach, since most of the area on that plane cannot be used for waveguide integration.

To date, much of this research has focused specifically on crossing loss mitigation, and crosstalk is generally avoided with the assumption of perpendicular (or significantly angled) waveguide orientations at overlapped regions on the wafer, to limit evanescent coupling-induced crosstalk.  Such a routing/layout scheme inherently has poor utilization of the available surface area, and is incompatible with conventional, Manhattan-type routing layouts in which nearby paths will lie parallel to each other for considerable distances.  An interconnect layout that prohibits co-propagation of out-of-plane waveguides will also increase the number of crossings, and thus increase the optical loss.

\begin{figure*}[]
	\centering
	\fbox{\includegraphics[width=\linewidth]{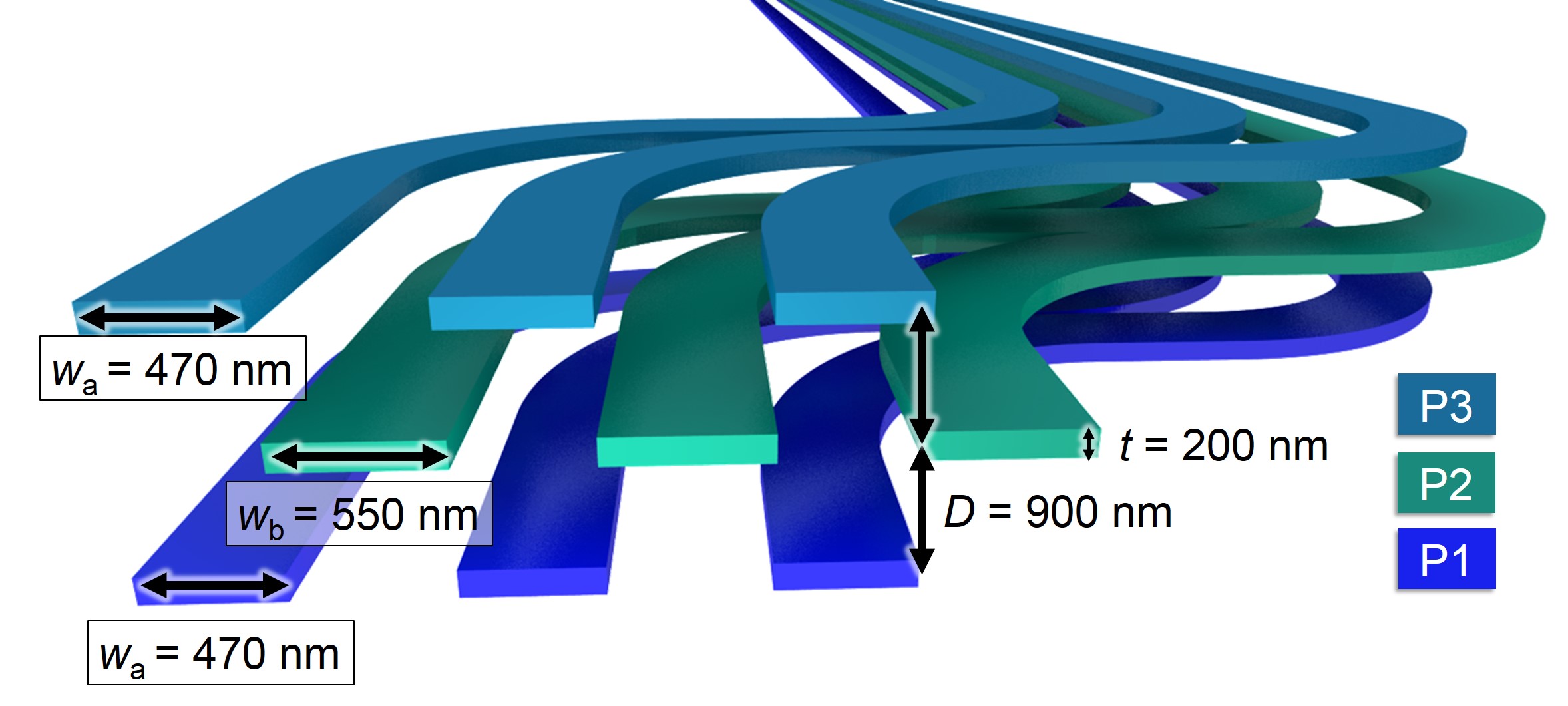}}
	\caption{Proposed 3D integration platform, utilizing alternating waveguide widths to mitigate crosstalk, which increases the waveguide packing density and allows arbitrary co-propagation lengths.}
	\label{fig:proposed}
\end{figure*}

The ideal 3D photonic integration architecture allows fully-packed waveguide integration (density-limited by lateral coupling) on \textit{each} additional plane, allows Manhattan-style routing with both perpendicular and parallel paths for different planes, and realizes compact, low-loss crossings and transitions, allowing maximum flexibility to the routing layout - a crucial consideration for further scaling.  To realize these goals, we propose a 3D integration strategy comprising an efficient IPC design and a robust optical routing technique. We experimentally demonstrate the system's performance in the key performance metrics of crossing loss, crosstalk, and interplane coupling loss.  Additionally, to assess the film properties of the stack, we fabricate and characterize microring resonators on each of the three planes.  The proposed platform is represented in Fig. \ref{fig:proposed}. It employs 200 nm-thick a-Si waveguiding planes with an interplane pitch of 900 nm.  For each a-Si plane in the stack, the nominal width of routed waveguides is alternated between two values, $w_{a} = 470$ nm and $w_{b} = 550 $ nm.  In this way, continuous constructive interference between adjacent planes is prevented via a phase mismatch, allowing these waveguides to be co-propagated over arbitrary distances; this is analogous to the use of superlattices for increasing the horizontal packing density of a single plane of waveguides \cite{song2015high}.  In our proposed scheme, waveguides on different planes are also staggered \cite{zhang2015multilayer} with a horizontal offset (half of the intraplane waveguide pitch) to further limit crosstalk without compromising the packing density.  In effect, these choices allow a smaller interplane pitch and relax the demands on the IPC.  

\section{\label{sec:fab}FABRICATION}

\begin{figure*}[]
	\centering
	\fbox{\includegraphics[width=\linewidth]{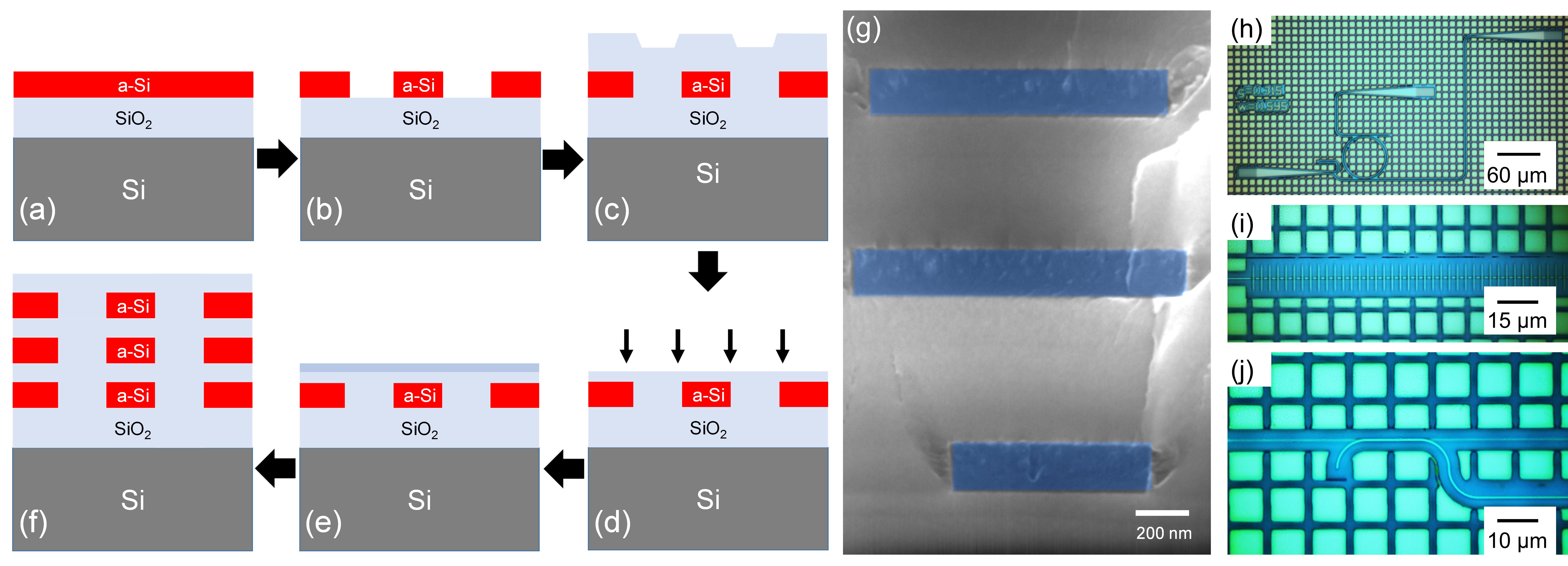}}
	\caption{(a-f) Fabrication flow for this work; (a) Deposition of first a-Si film; (b) Patterning/etching waveguide structures; (c) Deposition of spacer SiO${_2}$ plane (800 nm); (d) Chemical-mechanical planarization (CMP) of spacer ($\sim$300 nm depth); (e) Deposition of a $\sim$200 nm SiO$_{2}$ thickness compensation plane to reach target gap; (f) Repeating (a-e) based on number of desired planes.  Note that the top surface was not CMP'd after the final cladding deposition of 1100 nm.  (g) Scanning-electron-microscope cross-section of a-Si fill patterns in three planes; (h-j) Optical micrographs of representative test devices on the wafer.}
	\label{fig:process}
\end{figure*}

The proposed platform was prototyped at the Boulder Microfabrication Facility at NIST.  The fabrication flow is detailed in Fig. \ref{fig:process}(a-f).  For this study, three waveguiding planes denoted P1, P2 and P3 were employed, though the process is in principle scalable to larger numbers.  The a-Si deposition was performed with an inductively-coupled plasma chemical-vapor-deposition (ICP-CVD) system, utilizing SiH$_4$ / Ar chemistry at 150 $^{\circ}$C.  Prism-coupling measurements indicate a refractive index value of $3.12 \pm 0.1 $ and a slab propagation loss in 144 nm-thick films of $\sim 1.4$ dB per cm at $\lambda = 1550$ nm.  Patterning was performed with electron-beam lithography. ICP reactive-ion etching utilized a SF$_6$ / C$_4$F$_8$ chemistry.  Unused areas were patterned with a periodic partial fill to homogenize the surface and limit film stress.  Images of the finished sample after fabrication are shown in Fig. \ref{fig:process}(g-j).  The experimental interplane pitch of $\sim$700 nm is smaller than the design value of 900 nm, which may be explained by inaccuracies from using a white-light interferometer to track film thicknesses throughout the fabrication.

\section{\label{sec:char}CHARACTERIZATION}

The fabricated devices were characterized via a tunable laser source and detector system.   Light was coupled on and off-chip via fully-etched grating couplers and single-mode fibers at a nominal wavelength of 1540 nm.  In the data presented in the following sections, statistical uncertainties are reported as the standard deviation in transmitted optical power for sets of reference paths consisting of two grating couplers and a waveguide.
\subsection{\label{sec:rings}Microring resonators}

\begin{figure*}[t]
	\centering
	\fbox{\includegraphics[width=\linewidth]{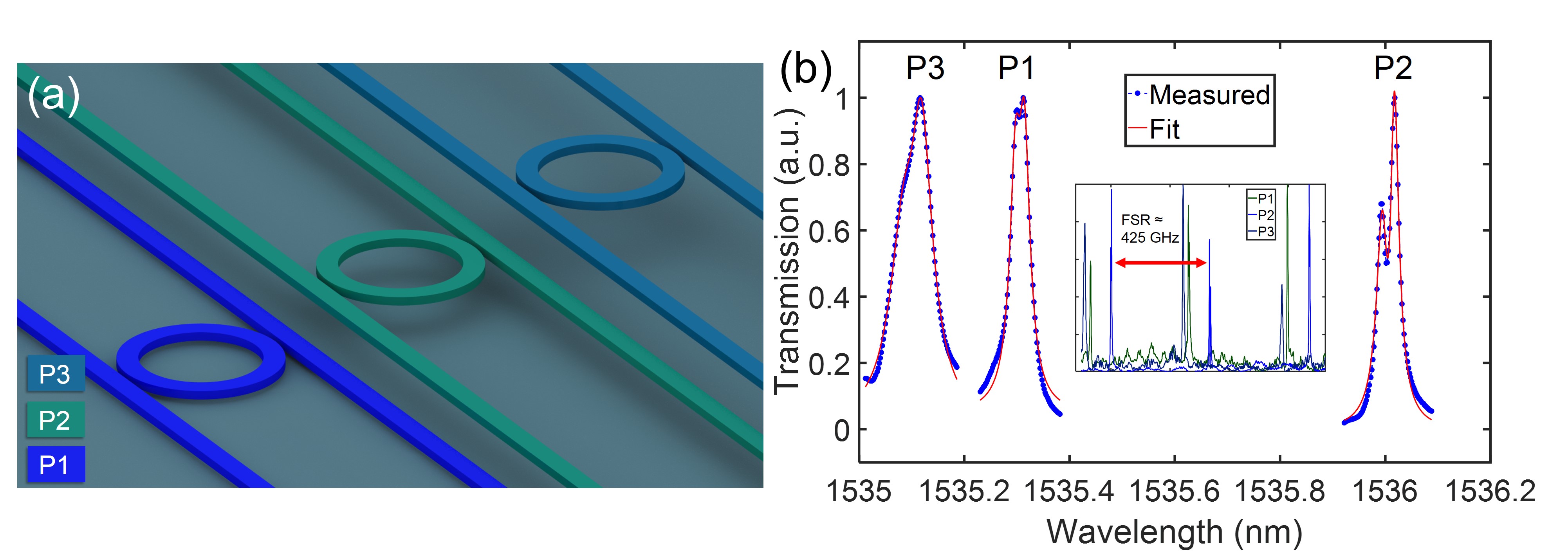}}
	\caption{Three planes of micro-ring resonators; (a) Test device arrangement; (b) Measured drop-port transmission spectra of one resonance from each ring, showing both raw and fitted curves.  Inset: Drop-port transmission spectra of identical P1, P2 and P3 rings encompassing three resonance peaks, showing a nominal free-spectral range (FSR) of $\sim$425 GHz.}
	\label{fig:rings}
\end{figure*}

The waveguiding performance and material quality of each of the three planes (P1-P3) was assessed by fabricating and measuring microring resonators with radii of 30 $\mu$m (Fig \ref{fig:rings}).  A ring-bus coupling gap of 500 nm was employed, incurring minimal loading.  A set of grating couplers (input, output, and drop ports) was fabricated with each ring.  The measured and normalized drop-port transmission for one doublet resonance from each ring (waveguide width of 550 nm for all three planes) is plotted in (Fig \ref{fig:rings}(b)), as well as the fitted value based on coupled-mode theory \cite{popovic2008theory}.  The loaded quality factors (Qs) for the P1,P2 and P3 doublet pairs (with the two peaks in the doublet denoted \textit{a} and \textit{b}) are as follows: P1 - $Q_{a} = 6.1 \times 10^4, Q_{b} = 6.4 \times 10^4 $; P2 - $Q_{a} = 6.2 \times 10^4, Q_{b} = 8.2 \times 10^4 $; P3 - $Q_{a} = 2.5 \times 10^4, Q_{b} = 3.2 \times 10^4 $.  These values are likely predominantly limited by pattern and etch-induced sidewall roughness (based on the earlier observed slab propagation loss of 1.4 dB per cm), which was not optimized in this work.

\subsection{\label{sec:ipcs}Interplane couplers}

\begin{figure*}
	\centering
\fbox{\includegraphics[width=\linewidth]{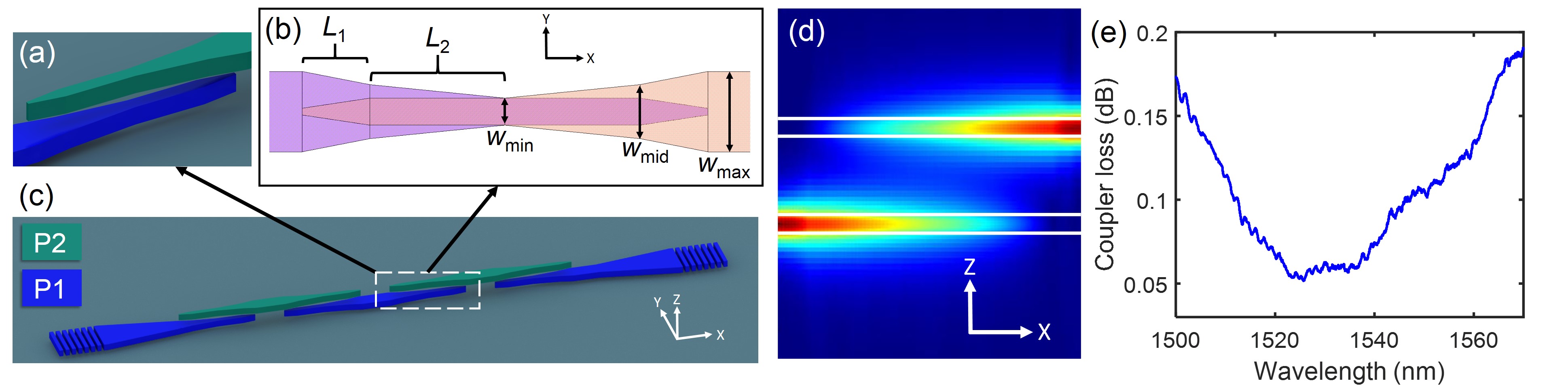}}
\caption{Efficient and compact interplane coupler design; (a) Zoom view; (b) Layout schematic showing key design parameters; (c) Perspective view of simplified layout for cut-back measurements; (d) 2D E-field slice of simulation result, showing complete power transfer; (e) Experimentally measured loss spectrum of 32 successive IPCs.}
\label{fig:ilc1}
\end{figure*}

Next, we consider the IPC design and characterization.  The designed interplane pitch of 900 nm, combined with the high-confinement a-Si core, poses a challenge for the IPC.  State-of-the-art IPCs for similar interplane pitches exhibit typical lengths between 100-200 $\mu$m \cite{Shang2015,Itoh2016}, or compromise the efficiency for shorter device lengths ($\sim$1 dB over a 60 $\mu $m long coupler\cite{Li2014}).    

An effective IPC design, consisting of a tapered width transition between two waveguides, should behave adiabatically (which enhances bandwidth and tolerance to fabrication errors), but should also be designed to enhance the evanescent coupling strength between the two waveguides.  This can be achieved with narrower waveguides to reduce the mode confinement.  For large interplane pitches, the average waveguide width throughout the transition should be minimized to the point that it does not introduce losses due to sidewall roughness.  However, a simple linear taper of the waveguide width between the maximum and minimum values results in excessively long couplers, since little coupling occurs until the waveguide dimensions are significantly narrowed.  We have thus implemented a two-level IPC design, making use of a ``fast'' initial taper to rapidly compress the waveguide width at the outer regions, combined with a ``slow'' extended taper region over which a much smaller width transition occurs (Fig. \ref{fig:ilc1}(b)).  The result is strong coupling over most of the useful taper length, while eliminating unnecessary space for bulk width adjustments at the input/output.  Compared to a simple uniform directional coupler approach, this has increased tolerance to thickness variations between layers.  The proposed design has the parameters $P_{1} = 4$ $\mu$m, $P_{2} = 15 $ $\mu$m, $w_{min} = 320$ nm,  $w_{mid} = 350$ nm and  $w_{max} = 510$ nm, with a total length of 38 $\mu$m.  The simulated insertion loss is 0.032 dB per coupler at a wavelength of 1540 nm via 3D finite-difference time-domain (FDTD).  A series of parametric variations near these design parameters was fabricated.  Each design was tested in a cut-back arrangement by comparing the spectral transmission of 32 successive transitions between P1/P2 to the averaged spectral transmission of twelve reference waveguide paths of the same total length, distributed across the test array.  The resultant loss spectrum of the best-performing design observed is plotted in Fig. \ref{fig:ilc1}(e).  A minimum loss of 0.05 $\pm$ 0.02 dB per coupler is observed at a wavelength of 1526 nm.  A loss better than 0.1 dB per coupler is maintained over a 35 nm span from $\lambda = 1512$ nm to $1547$ nm.  The measured device has on-mask parameters $L_{1} = 3$ $\mu$m, $L_{2} = 15 $ $\mu$m, $w_{min} = 330$ nm,  $w_{mid} = 370$ nm and  $w_{max} = 510$ nm, comprising a total length of 36 $\mu$m.  The difference in optimal design parameters likely comes from the reduced interplane pitch in the fabricated structure, leading to stronger-than-expected coupling.
\begin{figure*}[t]
	\centering
	\fbox{\includegraphics[width=\linewidth]{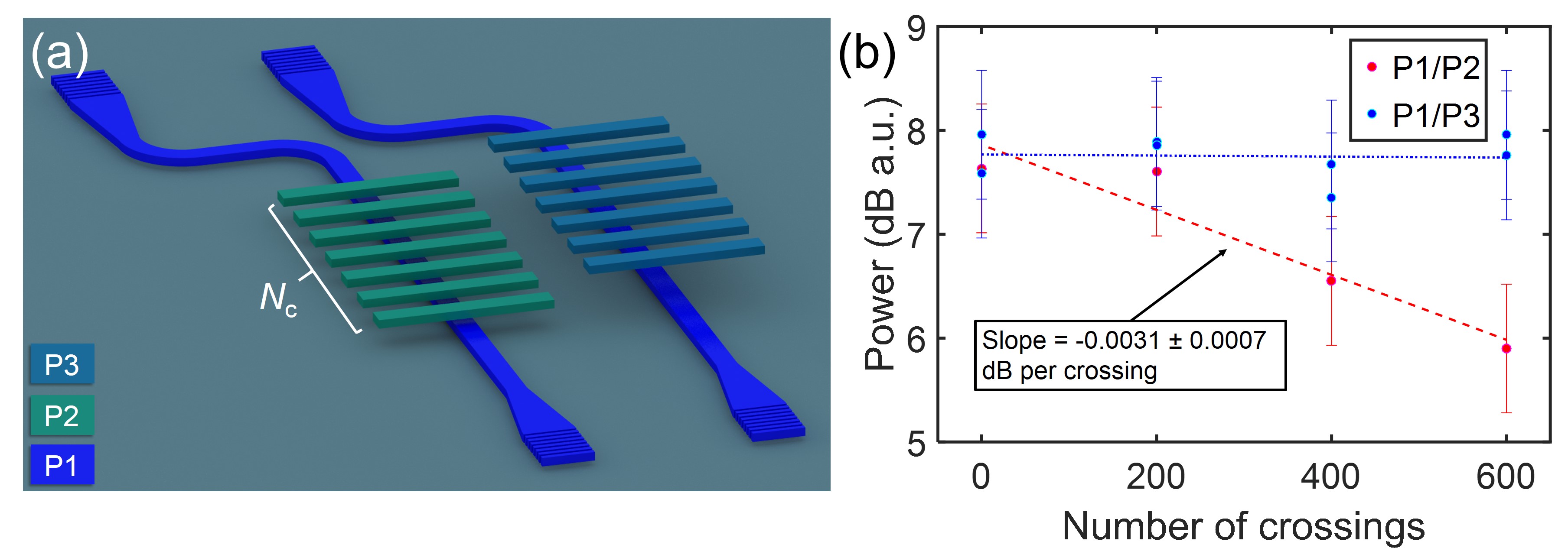}}
	\caption{Evaluation of waveguide crossing performance; (a) Simplified perspective view of test devices for P1/P2 and P1/P3 crossings; (b) Experimentally measured transmitted power in cut-back measurements.}
	\label{fig:crossing}
\end{figure*}

\subsection{\label{sec:crossings}Waveguide crossings}

Next, we investigate the performance of perpendicular out-of-plane waveguide crossings.  Test devices (Fig. \ref{fig:crossing}(a)) were fabricated with $N_{c}=$ 0, 200, 400 and 600 crossings for both P1/P2 and P1/P3 types.  The waveguide stubs acting as crossings were separated from each other by a pitch of 3 $\mu$m.  For P1/P3 crossings, the loss per crossing is below the measured standard error of $3\times10^{-4}$ dB per crossing.  For P1/P2 crossings, the measured value is $3.1\times10^{-3} \pm 7\times10^{-4}$ dB per crossing, on-par with the best measured to date\cite{sacher2016multilayer}, without the need for a dedicated plane to expand the interplane pitch.  These results demonstrate the scalability of this integration strategy to large waveguide packing densities.

\subsection{\label{sec:crosstalk}Crosstalk}

\begin{figure*}[t]
	\centering
	\fbox{\includegraphics[width=\linewidth]{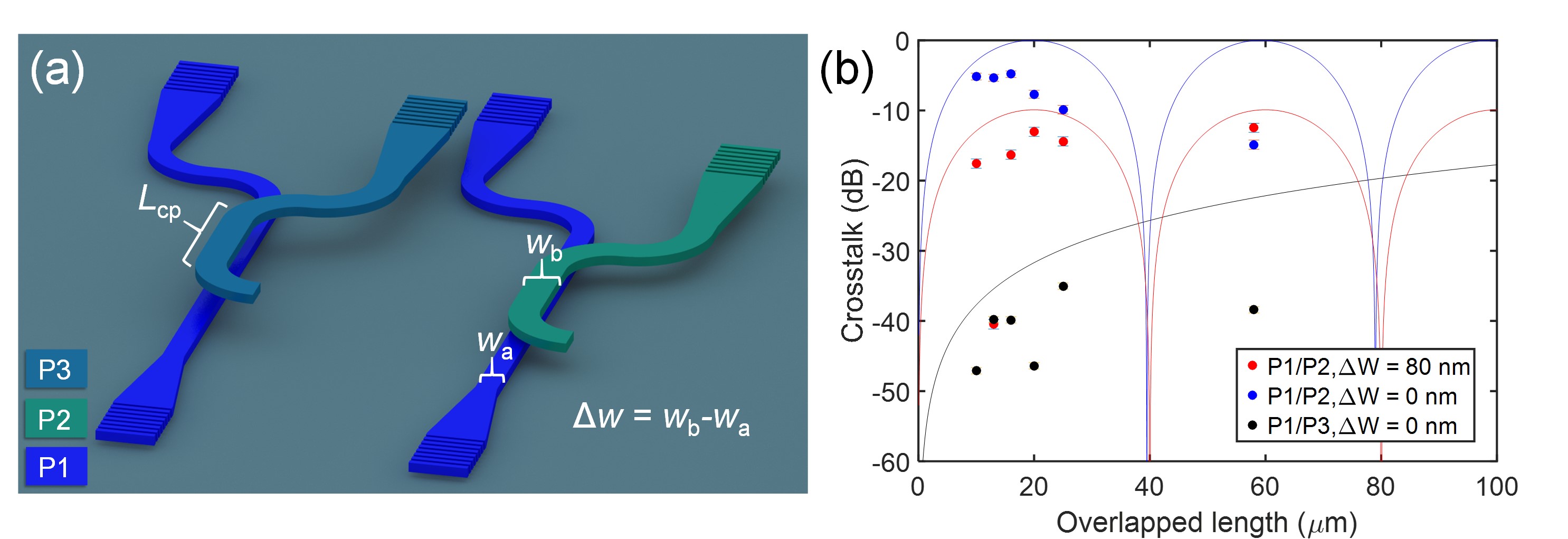}}
	\caption{Waveguide crosstalk evaluation; (a) Simplified perspective view of test devices for P1/P2 and P1/P3 crosstalk paths; (b) Experimentally measured crosstalk values (dots) and theoretical crosstalk data (solid lines).}
	\label{fig:crosstalk}
\end{figure*}

In order to effectively utilize the space available, and to avoid dilemmas in the routing, the crosstalk between co-propagating waveguides on different planes must be managed.  We now explore the performance of phase velocity mapping of waveguides on adjacent planes via a small difference in waveguide width.  This was done by co-propagating P1 and P2 or P3 waveguides for a variable distance and measuring the maximum ratio of upper-waveguide power to the total power from both arms.  For P1/P2 devices, both cases of $\Delta w = 0$ nm and $\Delta w = 80 $ nm were considered, while the P1/P3 case utilized the same nominal waveguide widths.  Test devices (Fig. \ref{fig:crosstalk}(a)) were fabricated and measured, with the results plotted in (Fig. \ref{fig:crosstalk}(b)).  The theoretically predicted crosstalk behavior for the fabricated test structures (via FDTD simulations) is also plotted to provide a comparison.  The highest values of crosstalk occur at different lengths due to differences in propagation constant and coupling strengths in each case.  For overlapped P1/P2 waveguides with identical widths, a severe maximum crosstalk of $-4.8 \pm 0.7$ dB was measured for a co-propagation length of 16 $\mu$m.  However, using a difference of 80 nm in the waveguide width, the crosstalk was dramatically reduced to $-12.5 \pm 0.7$ dB (58 $\mu$m co-propagation length), even in the extreme case of direct overlapping.  FDTD simulations show that a straightforward crosstalk improvement (based on the observed performance so far) to $<$ -33 dB is achievable for P1/P2 phase-velocity-mapped waveguides by offsetting them by 1 $\mu $m in the horizontal direction when co-propagation is required (see Fig. \ref{fig:proposed}). This would have no significant impact on the available surface area, since the same intraplane pitch can still be used.  Finally, for the P1/P3 overlapped case, a negligible crosstalk value of $-35 \pm 0.7$ dB was experimentally observed (at 25 $\mu$m length).  At the maximum measured length of 58 $\mu$m, the experimentally observed P1/P3 crosstalk is 16 dB smaller than the theoretical value.  This is most likely due a minor difference in thickness between the P2/P3 a-Si films, resulting in a coherence length much shorter than that of maximum coupling.

\section{\label{sec:conclusion}CONCLUSION}

In conclusion, we proposed a strategy for efficient photonic routing in 3D-integrated systems.  A prototype implementation was experimentally realized with three planes of amorphous silicon waveguides. Detailed characterization reveals exceptional performance in the critical performance metrics of out-of-plane crossing loss, interplane coupler loss, and crosstalk. Microring resonators were fabricated on all three planes, showing a quality factor up to $8.2 \times 10^{4}$.  An out-of-plane waveguide crossing loss of $3.1\times10^{-3} \pm 7\times10^{-4}$ dB per crossing for adjacent planes (P1/P2) was observed, and for double-spaced planes (P1/P3), the crossing loss was below the measurement limit of $3\times10^{-4}$ dB per crossing.  The large interplane pitch was bridged with a compact and efficient two-stage interplane coupler (IPC) design, showing a peak performance of $0.05 \pm 0.02$ dB per coupler at $\lambda = 1526$ nm.  Next, anticipating that Manhattan-style routing will be a necessary feature of high-density 3D optical interconnects, we investigated a means of enabling waveguides on adjacent planes to be propagated parallel to each other for arbitrary distances, without introducing excessive crosstalk.  By slightly modifying the waveguides on alternate planes to be 80 nm wider, continuous constructive interference is disrupted.  Directly-overlapped waveguides employing this technique showed a nearly sixfold reduction in crosstalk compared to those with identical widths.  This could later be combined with a simple constant horizontal offset (half of the intraplane pitch) that will lead to $<$-33 dB crosstalk between P1/P2 waveguides.  These results, showing drastically increased layout flexibility and space-efficiency, bolster the case for 3D integrated photonics.

We thank A. Rao for performing the prism coupling measurements and A. N. McCaughan for assistance with the layout software.  This is a contribution of NIST, an agency of the US government, not subject to copyright.

\nocite{*}
\bibliography{multilayer}

\end{document}